\documentstyle[preprint,aps,epsf]{revtex}

\newcommand{\referencestyle}{
\small
\abovedisplayskip=6pt
\belowdisplayskip=6pt
\vspace{12pt}}

\def\De{\Delta\eta}
\def\Det{\Delta\eta_T}
\def\Des{\Delta\eta_*}
\def\Dt{\Delta\tau}
\def\Dy{\Delta y}
\def\t0{\tau_0}
\def\y{\mbox{\rm y}}
\def\ch{\cosh}
\def\sh{\sinh}
\def\ben{\begin{eqnarray}}
\def\enn{\end{eqnarray}}
\def\ov{\over\displaystyle\strut}
\def\l({\left(}
\def\r){\right)}
\def\dst{\displaystyle\strut}

\begin{document}

\rightline{LUNFD6/(NFFL-7081)-Rev. 1994}
\rightline{hep-ph/9409327}
\begin{center}
{\large \bf Bose-Einstein Correlations\\
for Longitudinally Expanding, Finite Systems
}
\end{center}

\begin{center}
{\rm T. Cs\"org\H o$^{1,2}$\footnote{\rm E-mail: csorgo@sunserv.kfki.hu }
}
\end{center}
\begin{center}
{\it
$^1$KFKI Research Institute for
Particle and Nuclear Physics of the \\
Hungarian Academy of Sciences,
H--1525 Budapest 114, P.O. Box 49. Hungary \\
$^{2}$Department of Elementary Particle Physics, Physics Institute, \\
University of Lund,
S\"olvegatan 14,
S - 223 62 Lund, Sweden
}
\end{center}
\begin{center}
{August 17, 1994}
\end{center}
\vfill

\begin{abstract}
Bose-Einstein correlations and momentum distributions are
calculated for a longitudinally expanding boson source,
where the expanding system has a finite size in space-time
rapidity.
Such systems are physically realized in any of the single
jets in high energy $e^+e^-$, lepton-hadron and hadron-hadron
collisions as well as in high energy heavy ion collisions,
where the projectile is not really heavy.
 The 1D expansion generates a thermal length-scale
in the longitudinal direction, which
together with the finite length of the expanding tube
effects both
the momentum distribution and the Bose-Einstein
correlation function.
The Bose-Einstein correlations are shown to be more sensitive to
the smaller,
while the momentum distribution to the longer
of the two longitudinal length-scales.
\end{abstract}

\vfill\eject

{\it Introduction.} There is a renewed interest in the study of
relativistically
expanding one dimensional systems partly due to the recent high
energy heavy ion programme at CERN and AGS.
The underlining space-time picture is thought to be determined by
the Bose-Einstein correlation measurements. Both the Bose-Einstein
correlations and the momentum-spectra were determined,
 see e. g. the contributions of
the NA35 and NA44 collaborations in ref.~\cite{QM}.
In case of high energy elementary particle reactions
where jets dominate the underlining space-time picture,
usually only static parameterizations of the Bose-Einstein
correlation functions (BECF-s) are used, with very few
exceptions. For an introduction and review see ref.~\cite{bengt}.

In this Letter the importance of the dynamical picture is
emphasized. It seems to be necessary to analyze simultaneously
the invariant momentum distribution (IMD) as well as the
BECF-s in the {\it longitudinally
comoving system}, LCMS, in order to measure the usually very long
longitudinal size of the jets.

Particle interferometry for one dimensionally expanding
hydrodynamical systems
were studied first in ref.~\cite{sinyukov} where Bose-Einstein
interferometry was considered for an infinite, longitudinally
expanding Bjorken-tube.
Kolehmainen and Gyulassy considered the
one-dimensionally expanding ideal inside-outside
cascade~\cite{ka_gyu}, which was further elaborated in
refs.~\cite{pa_gyu}. The
correlations in between rapidity and space-time rapidity
 were considered for 1D expanding strings
in ref.~\cite{lutp} where also the relation of the
sidewards and outward momentum components were clarified.
However, in these works the rapidity distribution of the
final particles remained somewhat arbitrary and there was
only one characteristic length-scale in the longitudinal
direction. This characteristic length-scale,
the "length of homogeneity" or thermal length, appeared due to
the interplay of the flow-gradient and the temperature
in ref.~\cite{sinyukov} while it appeared as a characteristic
rapidity -- space-time rapidity correlation length in ref~\cite{lutp}.
In this Letter the interplay between the finite size of the
boson-emitting source together with the finite "thermal length"
shall be considered in detail.

One-dimensionally expanding systems are thought to be realized
 by the hadronic strings in the case of 2-jet
events in high energy elementary particle reactions.
Any of the jets of high energy $pp, p\bar p, e^+e^-$
or other $l + h$ collisions corresponds
to an approximately one-dimensionally expanding system.
Boson interferometry for decaying hadronic strings was
studied first in ref.~\cite{ander} and later in details
in ref.~\cite{bowler}.
In these publications, however,
the interplay between the total longitudinal length
of the string and the BECF
was not considered.
 High energy heavy ion collisions
may also create one-dimensionally
expanding systems, especially in the case of light projectiles.
Heavier projectiles may create three-dimensionally
expanding systems which shall not be
discussed here.

Both the momentum spectra and
the BECF are prescribed
in the applied Wigner-function formalism~\cite{pratt_csorgo,bengt}.
In this formalism the BECF is calculated from the two-body Wigner-function
assuming chaotic particle emission. In the final expression the
time-derivative of the Wigner function
is approximated~\cite{pratt_csorgo,bengt}
 by a classical emission function $S(x;p)$,
which is the probability that a boson is produced
at a given $ x = (t, \vec r \,) = (t,x,y,z)$ point in space-time
with the four-momentum $p = (E, \vec p\,) = (E, p_x, p_y, p_z)$.
The particle is on the mass shell, $ m^2 = E^2 - \vec p^{\,\, 2} $,
and the off-shell emission functions entering into the
BECF~\cite{pratt_csorgo}
are approximated by the off-shell continuation of the
 on-shell emission  functions.
In this Letter the effect of final state  Coulomb and Yukawa
interactions
shall be neglected.

{\it The emission function.} $\,\,$
We model the emission function in terms of the longitudinally boost-invariant
variables:  the longitudinal proper-time,
$\tau = \sqrt{t^2 - z^2}$, the
space-time rapidity $\eta = 0.5 \ln\l({\displaystyle\strut t+z\ov t-z}\r)$,
the transverse mass $m_t = \sqrt{E^2 - p_z^2}$
 and the momentum-space rapidity $\y = 0.5 \ln\l({\displaystyle\strut
E + p_z \ov E - p_z}\r)$.
We shall assume that the emission function is characterized by a given
space-time rapidity distribution $G(\eta)$,
a distribution in the transverse production points $I(r_t)$
and a proper-time distribution $H(\tau)$. The correlations
between rapidity and space-time rapidity are taken into account
by the conditional probability $J_{m_t}(\eta - \y)$, which gives the
probability
that a boson is emitted with rapidity $\y$ from the point-like source
characterized by the space-time rapidity $\eta $.
For a thermal emission, the width of this function is a function of the
transverse mass $m_t$, as we shall see below. We may also introduce
an arbitrary $g(\vec p_t)$ transverse momentum
distribution. Thus the emission function can be written in the following way:
\ben
S(x;\y,\vec p_t) \,\, d^4 x & = &
	 g(\vec p_t)\,\, G(\eta)\,\,  H(\tau)\,\, I(r_t)\,\,
			J_{m_t}(\eta - \y) \,\, d\eta \,\, d\tau  \,\, d^2
				\vec r_t.
\enn
First, we shall consider results for  general, unspecified functional form
of the spacetime distribution functions $ G(\eta),\, H(\tau) $ and $I(r_t)$
and discuss how these quantities can be determined from the simultaneous
analysis of the BECF and that of the IMD.
We shall also apply  Gaussian approximations
for the distribution functions of $ \tau$, $\eta$, $\eta - \y$
 and $r_t$ as follows
\ben
        G(\eta) & =  & {\displaystyle\strut 1\ov (2 \pi \De^2)^{1/2} }\,
           \exp\l( - {\displaystyle\strut \eta^2 \ov 2 \De^2} \r) \label{e:g}\\
        I(r_t) & = & {\displaystyle\strut 1\ov (2 \pi R_G^2 )}\,
        \exp\l( - { \displaystyle\strut r_t^2 \ov 2 R_G^2 } \r) \label{e:i}\\
        H(\tau) & = & {\displaystyle\strut 1\ov (2 \pi \Dt^2 )^{1/2}} \,
          \exp\l( - {\displaystyle\strut (\tau - \t0)^2 \ov 2 \Dt^2} \r),
		\label{e:h}\\
	J_{m_t}(\eta - \y) & = &
	   {\displaystyle\strut 1\ov (2 \pi \Det^2)^{1/2} }\,
	\exp\l(\displaystyle\strut - {(\eta - \y)^2 \ov 2 \Det^2} \r).
		\label{e:det}
\enn
This Gaussian ansatz results in a simple  analytic
parameterization of the BECF as well as for the rapidity-dependent part
of the IMD.
The transverse momentum dependent
factor, $g(p_t)$ remains unspecified and cancels from the results.

It is to be emphasized that by introducing a finite width $\De$ we break
the boost-invariance of our source in the longitudinal direction too.
Thus we expect a non-stationary rapidity
distribution.
The stationary rapidity distribution corresponds to
the $\Delta \eta_T \rightarrow \infty $ limit.

In this model, the source at a given space-time -- rapidity $\eta $
is emitting bosons with the rapidity $ \y =\eta$, however with a
non-vanishing width
of the rapidity distribution of the emitted bosons.
 This width, $\Det$, can be determined by Monte-Carlo
simulations in the case of decaying strings e.g. in the Lund model
$\Det \approx 0.4$, refs.~ \cite{lund,lutp}.
It can be calculated in the framework of hydrodynamics
from the relativistic Boltzmann-distribution,
$
f(x;p) = C \exp( - {\displaystyle\strut p\cdot u(x) \ov T}),
$
where the freeze-out temperature is denoted by $T$ and the
four-velocity of the 1D expanding matter at space-time point $x$
is given by the scaling Bjorken expansion~\cite{bjorken} as
\ben
u(x) & = &(t/\tau,\, 0,\, 0,\, z/\tau) = (\ch (\eta),\, 0,\, 0,\, \sh(\eta) )
\enn
Since the four momentum $p$ can be also written as
$p = (m_t \ch(\y), \vec p_t, m_t \sh(\y))$,
the relativistic Boltzmann distribution for the 1D expanding systems
can be rewritten as
\ben
f(x;\y,p_t) & = & J_{m_t}(\eta - \y) =
	 C \exp\l( - {\displaystyle\strut  m_t \ch(\eta - \y) \ov T} \r)
	 \approx
	C' \exp\l( - {\displaystyle\strut m_t (\y - \eta)^2 \ov T} \r)
		\label{e:b}
\enn
which is the same form as in eq.~\ref{e:det}.
Thus in the case of the thermal emission, the width of the
rapidity distribution of the particles is a decreasing function of the
transverse mass,
\ben
\Det\,(m_t) & = & \sqrt{\displaystyle\strut T\ov m_t} .
\enn
This completes the specification of the model.

{\it Evaluation of the space-time integrals.}
We can evaluate the integrals in the LCMS, the longitudinally
comoving system.
Here we parameterize the mean momentum of the particles as
$K = (K_0,K_{T,O},0,0)$ i.e. the longitudinal
component of the mean momentum is vanishing. Since the mean momentum
is purely transverse to the $z$ axis (jet or beam axis), it is
pointing towards a direction which is named ~\cite{bertsch,lutp}
the out direction. This we chose to coincide with our $x$ axis
and index by $(T,O)$.
The $\y$ axis shall be called the {\it side} component,
indexed with $(T,S)$ ({\bf T}ransverse, {\bf S}ide ).
In the Laboratory system the LCMS moves with the
mean rapidity of the pair,
$ Y = {\displaystyle\strut 1 \ov 2}
\ln \l( {\dst E_1 + E_2 + p_{z,1} + p_{z,2} \ov
\dst E_1 + E_2 - p_{z,1} - p_{z,2}} \r) $.
In the LCMS, we index the variables with (').
We calculate the auxiliary function $\tilde S(\Delta k, K)$
in the LCMS for small values of the components of the relative momentum
$\Delta k = (\beta_T Q_{T,O},Q_{T,O},Q_{T,S},Q_L)$.
In the integration we utilize the following
equalities or approximations:

\ben
	\eta' \, = \, \eta - Y
	\qquad & \mbox{\rm and} & \qquad
	\y' \, = \, \y - Y \\
	t' \, = \, \tau\, \ch(\eta') \approx \tau
	\qquad & \mbox{\rm and} & \qquad
	z' \, = \, \tau\, \sh(\eta') \approx  \tau \eta' \approx \t0 \eta'.
\enn
 The latter two approximations are
valid when the
width of the rapidity distribution of the bosons emitted at a
fixed value of $\eta'$ is smaller than unity, $\Det < 1$, and when the width
of the proper-time distribution is smaller than the mean proper-time,
$ \Dt < \t0 $.
The first approximation is valid
for thermal emission in the $m_t $ region
$T < m_t$ as well as
e.g for the Lund strings ($\Det \approx 0.4$) .
        (Note, that in case of a decaying hadronic string, the fluctuations
        of the break-up points create the rapidity -- space-time rapidity
        correlation length $\Det$.)

 The second approximation is more safe for the case of
heavy ion collisions, where the mean freeze-out time is about $\t0 \approx 4-5$
 fm/c,
and the estimated duration is very short, $\Dt < 2$ fm/c,
{}~\cite{QM}.
With the help of the above approximations
the integrals can be evaluated.

{\it General results.}
The one-particle IMD can be calculated
for arbitrary space-time distribution functions as
under the assumption that the relativistic Boltzmann
distribution creates the correlations in between rapidity and space-time
rapidity, eq.~\ref{e:b}.

\ben
{\displaystyle\strut d^2n\ov d\y dm_t^2} & = &
	g(p_t) \, \, \int_{-\infty}^{\infty} d\eta \,\,
	 G(\eta) \, J_{m_t}(\eta - \y),
	\label{e:dng} \\
G(\y) & = & \lim_{m_t \rightarrow \infty} \,  {\dst 1 \ov g(\vec p_t) } \,
	{\dst d^2 n\ov d\y dm_t^2}, \label{e:gg}
\enn
Based on the above formulas, the function $G(\eta)$ can be measured
as the asymptotic (large $m_t$) limit of the rapidity
dependent part of the IMD $ d^2 / dy / dm_t^2$.
Then the measured $G(\eta)$
 function can be convoluted with the relativistic
Boltzmann distribution according to equations~\ref{e:b}, \ref{e:dng},
\ref{e:gg}.
Thus the $m_t$ dependent width in rapidity
 of the IMD is prescribed
from the model and can be contrasted to the data for a general
shape of the $G(\eta) $ distribution.
This  rapidity dependent part of the
IMD is obtained as a thermally broadened
space-time rapidity distribution, and the asymptotic, large $m_t$
limit of the rapidity distribution corresponds to the space-time
rapidity distribution. With decreasing $m_t$, the thermal broadening
becomes larger and larger except the cases when {\it i,} the
rapidity plateau becomes infinitely long,
or when {ii,} the decreasing
part of the rapidity distribution falls outside the experimental
acceptance.

The general result for BECF is
given in the LCMS of the particle pairs as
\ben
C(\Delta k, K)& =&  1 +
		\mid \tilde H(\beta_T Q_{T,O} ) \mid^2 \,
		\mid \tilde I(Q_{T,O},Q_{T,S} ) \mid^2 \,
		\mid \tilde G^{eff}_{Y,M_t}(\tau_0 Q_L) \mid^2,
\enn
Here $\tilde H$ etc. denotes the Fourier-transformed space-time
distribution and
the effective space-time rapidity region where the
boson pairs are emitted from is denoted by
\ben
G^{eff}_{Y,M_t} (\eta ) & = &
		{\dst 1 \ov \int d\eta \,\, G(\eta) \, J_{M_t} (\eta - Y) }
	 	\,	G(\eta) J_{M_t} (\eta - Y)
\enn
and the mean transverse mass is denoted by $M_t = (m_{t,1} + m_{t,2}) / 2 $.
Thus in the one dimensionally expanding case
the interpretation of the components
 of the correlation function is the following:
the duration in proper-time gives contribution to the out component
of the correlation function while the transverse distribution of the emission
points, $I(\vec r_t)$ yields a symmetric contribution to both the out and the
side component. The longitudinal component measures
that region in the longitudinal direction where
bosons with similar momenta may emerge from.

For an arbitrary distribution of rapidities, e.g. a flat central plateau
plus some wings, these results give an opportunity
for a measurement of the underlying, non-Gaussian space-time distribution.

{\it Results for the Gaussian ansatz.}
The general results may be further simplified in the framework
of the Gaussian approximations, \ref{e:g}-\ref{e:b} as follows:
\ben
{\displaystyle\strut d^2 n\ov dy dm_t^2 }
          & = & {\displaystyle\strut 1 \ov (2 \pi \Dy^2(m_t)\,)^{1/2}}
		\,\, g(p_t)\,\,
	 \exp( - {\displaystyle\strut y^2 \ov 2 \Dy^2(m_t)})
\enn
and
\ben
C(\Delta k, K)& =&  1 + \exp\l( - (R_{T,S}^2 Q_{T,S}^2 +
				R_{T,O}^2 Q_{T,O}^2 +
				R_L^2 Q_L^2) \r)
\enn

where the parameters are related to those of the source by
\ben
	\Dy^2(m_t) & = & \De^2 + \Det^2(m_t), \\
	R_{T,S}^2 & = & R_G^2, \\
	R_{T,O}^2 & = & R_G^2 + \beta_T^2 \Dt^2, \\
	R_L^2(m_t) & = & \t0^2 \Des^2(m_t) .
\enn
	The quantity $\Des$ corresponds to the effective
	correlation length between rapidity and space-time rapidity, as seen by the
	BECF. Its square is half of
	the harmonic mean of the
	squares of the thermal correlation length
	 and the total size of the pseudo-rapidity
	distribution,
\ben
	{\displaystyle\strut 1 \ov \Des^2(m_t) }& = &
	{\displaystyle\strut 1 \ov \De^2 }
		+ {\displaystyle\strut 1 \ov \Det^2(m_t) },
\enn
	thus it is always dominated by the {\it shorter} of the thermal
	and geometrical length scales ($\Det$ and $\De$, respectively).
	However, the square of the width of the rapidity distribution,
	$\Dy^2(m_t)$ is the quadratic sum of the geometrical and the thermal
	length scales, thus it is dominated by the {\it longer} of the two.

	{\it Discussion}.
	For thermally emitting sources the Boltzmann distribution
	predicts a very specific change of the rapidity-width
	of the  $d^2 / dy / dm^2_t $ distribution
	as a function of the transverse mass:

	\ben
		\Dy^2(m_t) & = & \De^2 + {\displaystyle\strut T\ov m_t}.
	\enn

	This effect, based on simple kinematics,
	can be experimentally checked by measuring the $\y$- width of the
	IMD-s at  different values of fixed $m_t$.
	Fitting the result with the above equation, both the geometrical
	size of the emission in pseudo-rapidity, $\De $
	 and the freeze-out temperature $T$
	can be extracted.

	The transverse mass dependence
	of the width of the rapidity distribution
	becomes especially well measurable in the intermediate region, where
	the thermal and geometrical length-scales are similar, see Fig. 1 and
	Fig. 2. as examples.

	In principle, it is possible to determine the same set of parameters
	from a measurement of the $m_t$ dependence of the
	$R_L$ parameter of the BECF-s as follows:
\ben
	R^2_L (m_t) & = & \t0^2 \,\, {\displaystyle\strut T \ov m_t} \,\,\,
				{\displaystyle\strut 1 \ov \displaystyle\strut
			1 + {\displaystyle\strut T  \ov m_t \De^2} }.
\enn

	This provides a possibility for the consistency check when the model is
	compared with the data. Note that this expression coincides with
	Sinyukov's formula for $\De \rightarrow \infty $,
	as it is expected since
	in that limit our model source corresponds to the infinite expanding
	Bjorken-tube. Thus the above expression can be
	 considered as the generalization
	of Sinyukov's formula for finite, 1D expanding systems.
	At present energies, we expect that the corrections are significant,
	The effects of finite geometrical size on the longitudinal radius
	parameter are visualized on Fig.3. and Fig.4. In both cases the
	mean freeze-out time was chosen to be $\t0 = 7.0$ fm/c. For $\De = 0.5$
	the correction at low $m_t$ yields about a factor of $0.5$ decrease
	in the longitudinal radius,
	because the finite geometrical size is comparable with the thermal
	length-scale for moderate values of $m_t$ in the case of
	Fig. 3. For a larger system with
	$\De = 1.0$ the finite size correction factor is about $0.8$.
	The difference between Sinyukov's formula and
	its finite-size corrected version vanishes for high
	values of the transverse mass and also decreases for increasing values
	of the geometrical size $\De$.

	The above expression is a function of three
	free parameters, $\t0, T$ and
	$\De$. To determine all the three parameters
	simultaneously from the $m_t$ dependence of the BECF
	one needs very dedicated set-up.
	It seems to be more straightforward to determine the $ T$ and $\De$
	parameters from the $m_t$ dependence of the rapidity - width
	of the $d^2 n/dy/dm_t^2$ distribution,
	and use the interferometry measurement  to infer the
	 value of the mean freeze-out time
	$\t0$.  Thus the importance
	of dedicated Bose-Einstein experiments and wide acceptance detectors
	is emphasized simultaneously: the
	BECF has to be studied as a function of the $m_t$
	and the $\y $ -width of the IMD
	has to be determined as a function
	of $m_t$, too.

	In the considered case, where the transverse position and the
	transverse momentum distribution decouple, the side component
	of the BECF measures the geometrical size, and the out component
	measures the quadratic sum of both the geometrical radius and the
	width of the proper-time distribution times the mean transverse
	velocity of the pair. The latter can be considered as an increase
	 of the effective source size in the out
	direction due to the evaporation
	and the mean speed of the bosons.

	{\it Limiting Cases.}
	 Depending on the relative size of the
	thermal and the geometrical logitudinal scales, $\De_T $ and $\De$,
	we have two limiting cases:
	{\it Case i,} If the geometrical size becomes much larger
	than the thermal length,
	 the rapidity-width of the IMD
	approaches the geometrical size and becomes independent of the
	transverse mass,

	\ben
	{\dst d^2 n \ov d\y dm_t^2 }
		& \propto & \exp( - {\dst y^2 \ov 2 \De^2})
	{\qquad \mbox{\rm if} \qquad } \De >> \De_T, \\
	R_L^2 & = & \tau_0^2 \De_T^2 = \tau_0^2 {\dst T \ov m_t}
	{\qquad \mbox{\rm if} \qquad } \De >> \De_T.
	\enn

	In this limit, corresponding to long and
	relatively cold expanding system, the {\it geometrical}
	longitudinal size is dominating the width of the
	{\it rapidity distribution}, while the {\it thermal}
	 longitudinal length-scale
	is dominating the longitudinal {\it radius
	parameter} of the BECF, measured in the LCMS of the boson pair.

	{\it Case ii,} If the
	rapidity distribution of
	the particles at a fixed space-time rapidity is
	much more wider than the geometrical length-scale,
	 then interferometry measures
	the geometrical size properly, and the temperature can be inferred
	from the width of the rapidity distribution:

	\ben
	{\dst d^2 n \ov d\y dm_t^2 }
		& \propto & \exp( - {\dst y^2 \ov 2 \De^2_T})
		  \propto \exp( - {\dst E \ov T} )
	{\qquad \mbox{\rm if} \qquad } \De_T >> \De, \\
	R_L^2 & = & \tau_0^2 \De
	{\qquad \mbox{\rm if} \qquad } \De >> \De_T.
	\enn

	This limiting case
	 corresponds to a relatively hot and small, expanding
	system. The effects of the expansion cancel from the results
	here. The expressions are the same as those for a hot
	source with no expansion.

Note that this work is an application of the ideas presented in~\cite{nr}
to the case of the relativistic one dimensionally expanding systems.

 {\it In summary}, we have calculated the rapidity distribution
and the Bose-Einstein correlation function in the longitudinally comoving
frame for
one-dimensionally expanding systems. In the longitudinal
direction, there are two length-scales present:
 the geometrical and the thermal one.
The latter is either due to the combined effect
 of the flow-gradient and the temperature
or e.g. to the stochastic nature of the string fragmentation.
In an analytically tractable model for the emission function,
the longitudinal component of the Bose-Einstein correlation function
is shown to measure dominantly the shorter while the
 width of the rapidity distribution
measures dominantly the longer length-scale.
This confers the intuitive picture that the strings are really long,
but cannot be seen by BEC measurements due to their expansion.
The geometrical size can be determined from the $m_t$ dependence
of the rapidity-width of the one-particle invariant
momentum distribution.
 The freeze-out temperature, the mean proper-time,
the width of the proper-time distribution as well as the geometrical
size in the transverse direction can be determined
within this picture by a combined use of the radius parameters in the
correlation function on one hand and by
the analysis of the transverse mass dependent
rapidity-width of the  invariant momentum distribution on the other hand.

{\it Acknowledgments:}
 I would like to thank B. L\"orstad for kind hospitality and stimulation
during my stay at University of Lund. I thank G. Gustafson and U. Heinz
for helpful discussions and thank to \'A. Till and L. P. Csernai  for kind
hospitality at the University of Bergen in March 1994.
 This work was supported in part by a NORFA grant,
by the NFR  and
the Hungarian Academy of Sciences exchange grant,
by the Human Capital and Mobility (COST) programme of the
European Economic  Community under grants No. CIPA - CT - 92 - 0418
(DG 12 HSMU), by the Hungarian
NSF  under Grant  No. OTKA-F4019, and also as a prelude
it is supported by the Hungarian - U. S. Joint Fund.

\vfill
\begin{figure}
          \begin{center}
          \leavevmode\epsfysize=5.0in
          \epsfbox{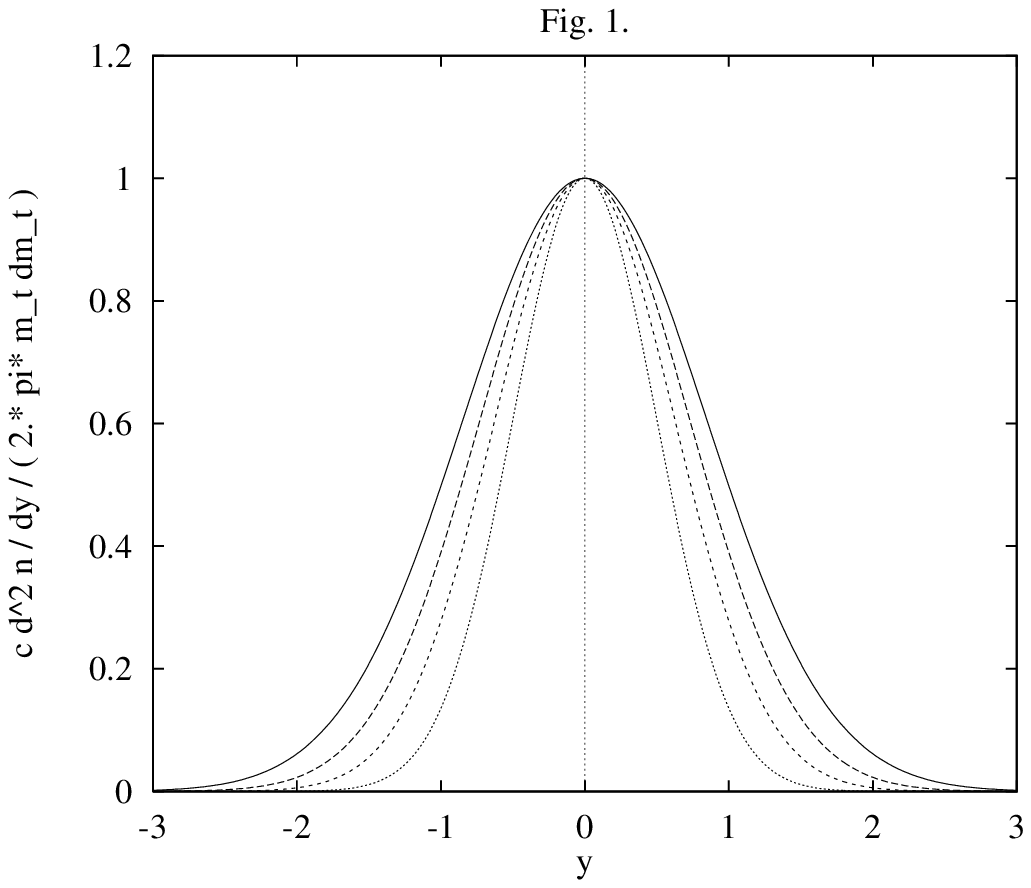}
          \end{center}
 \caption{The change of the rapidity-width of the
$d^2 n /dy / dm_t^2$  distribution for increasing
values of the (fixed) transverse mass,
$m_t = 0.3, 0.5, 1.0 $ GeV/c, as shown by the
solid, long-dashed and short-dashed
lines, respectively.
The dotted line stands for the asymptotic $m_t \rightarrow \infty$
limit, coinciding with $G(y)$, the geometrical shape of the source
in the longitudinal direction which was
chosen to be a Gaussian with $\Delta \eta = 0.5$.
The distributions were normalized to the same maximum value of 1.
 The pion mass ($m = 0.14$ GeV) and a freeze-out
temperature of $T = 0.14$ GeV was utilized. }
\end{figure}
\vfill\eject
\begin{figure}
          \begin{center}
          \leavevmode\epsfysize=5.0in
          \epsfbox{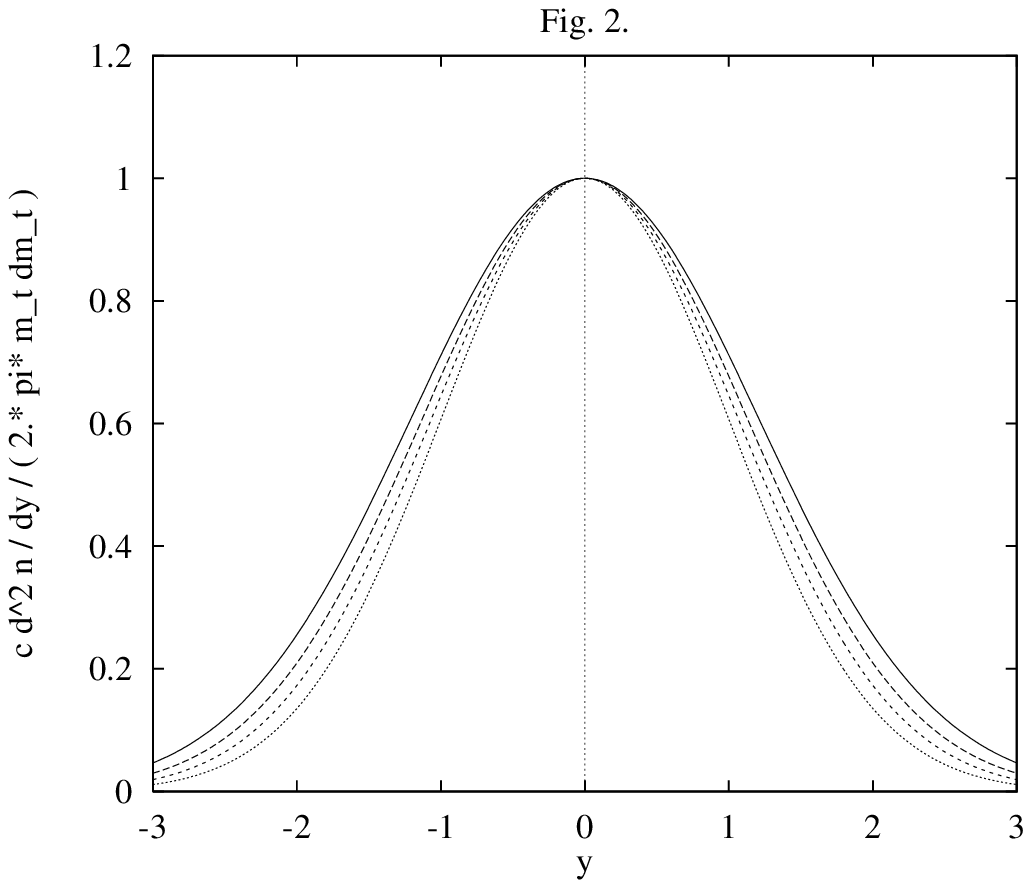}
          \end{center}
\caption{
Same as Fig. 1. but for $\Delta \eta = 1.0$ .
}
\end{figure}
\begin{figure}
          \begin{center}
          \leavevmode\epsfysize=5.0in
          \epsfbox{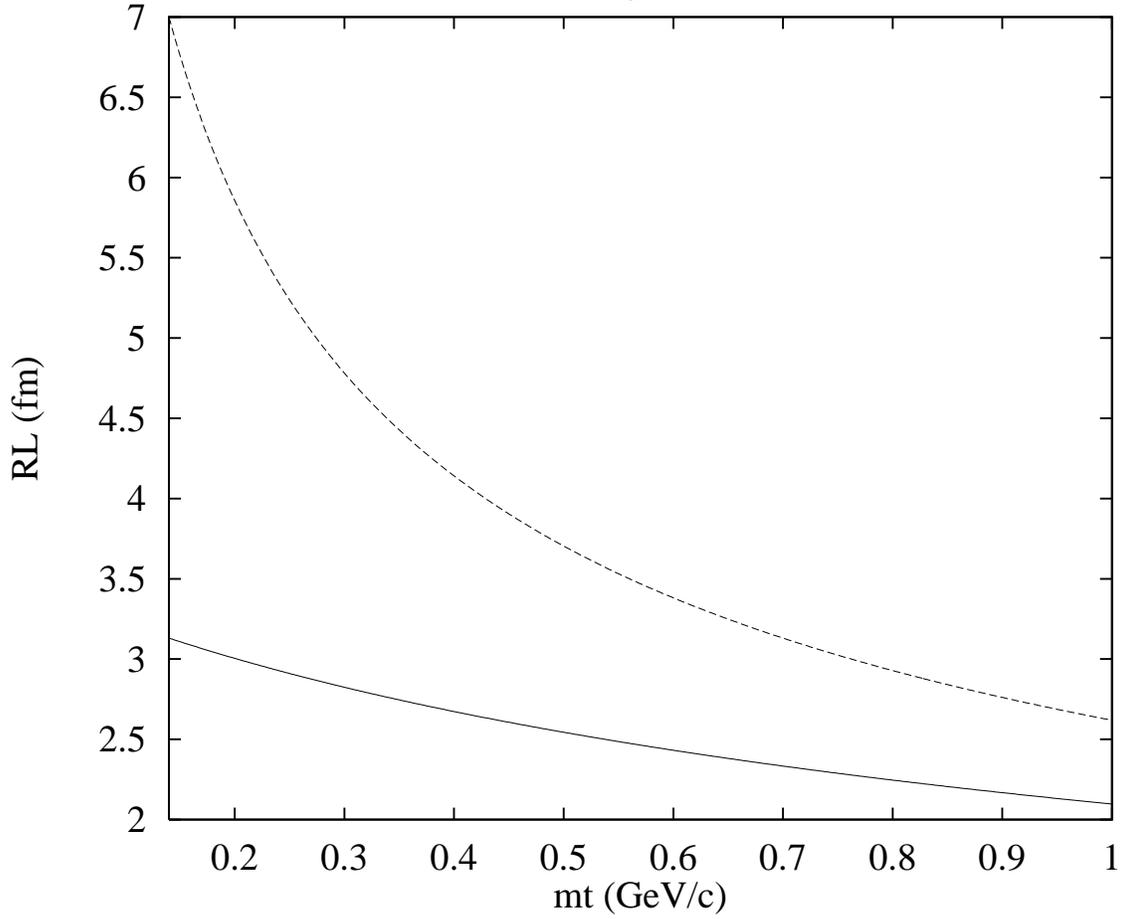}
          \end{center}
\caption{The $m_t$ dependence of the $R_L$ radius parameter for finite
1D expanding systems is shown by the solid curve.
The mean freeze-out time is chosen to be $t_0 = 7 $ fm/c and $\De = 0.5$.
The other parameters are the same as for Fig. 1. Dashed line shows
the prediction for $R_L$ based on Sinyukov's formula.
}
\end{figure}
\begin{figure}
          \begin{center}
          \leavevmode\epsfysize=5.0in
          \epsfbox{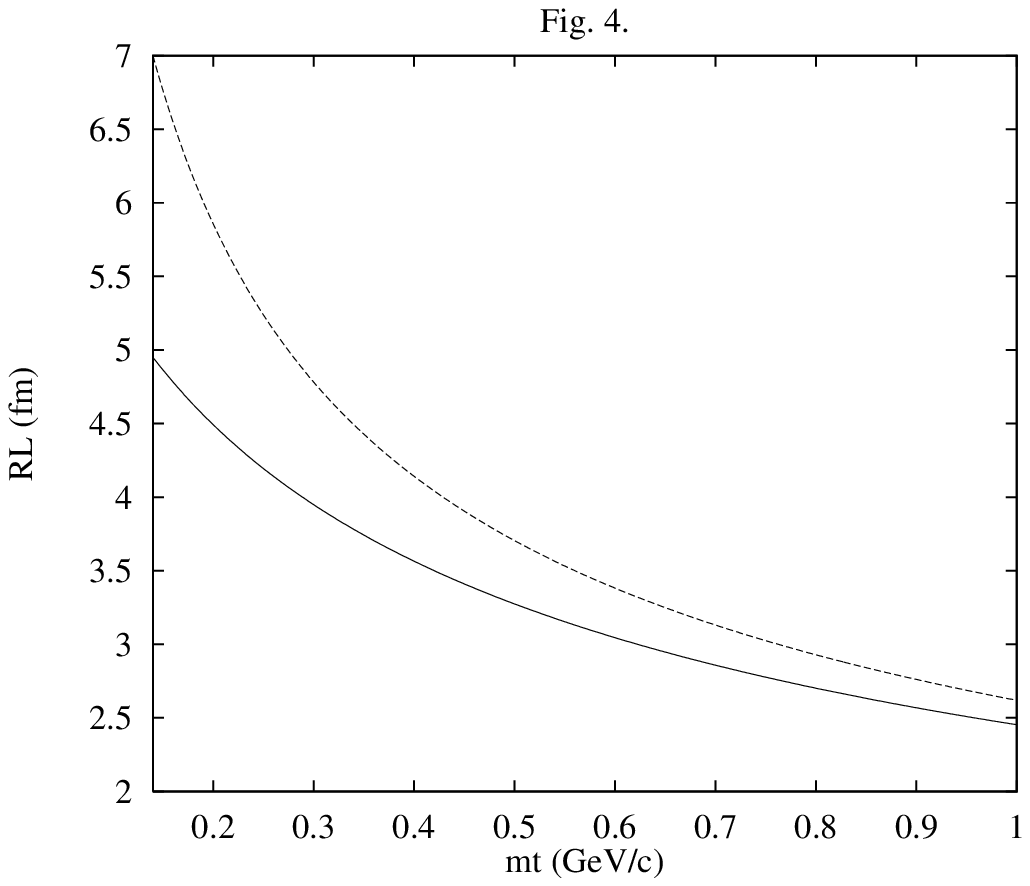}
          \end{center}
\caption{
Same as Fig. 3. but for $\Delta \eta = 1.0$ .
}
\end{figure}


\begin{thebibliography}{99}
\referencestyle
\bibitem{QM}            Proceedings of the Quark Matter conferences,
                        especially Nucl.\ Phys.\  {\bf A498},
                        (1989), Nucl.\ Phys.\ {\bf A525}, (1991),
                        Nucl.\ Phys.\ {\bf A544}, (1992) and
                        Nucl.\ Phys.\ {\bf A566}, (1993).
\bibitem{bengt}		B. L\"orstad,  Int. J. Mod. Phys.
        		{\bf A12} (1989) 2861-2896.
\bibitem{sinyukov}      A. Makhlin and Y. Sinyukov,
                        Z. Phys. {\bf C39}, 69 (1988).
        		(1986) 203.
\bibitem{ka_gyu} 	K. Kolehmainen, M. Gyulassy, Phys. Lett. {\bf B180}
\bibitem{pa_gyu} 	S. S. Padula and M. Gyulassy, Nucl. Phys.{\bf
        		B339} 378 (1990)
\bibitem{lutp}        	T. Cs\"org\H o and S. Pratt,
                        {\bf KFKI-1991-28/A}, p. 75.
\bibitem{ander}		B. Andersson and W. Hoffman, Phys. Lett. {\bf B169}
			(1986) 364
\bibitem{bowler}	M. G. Bowler, Particle World {\bf 2} (1991) 1.-6.
\bibitem{pratt_csorgo}	S. Pratt, T. Cs\"org\H o, J. Zim\'anyi, Phys. Rev.
        		{\bf C42} (1990) 2646;
                        S. Pratt, Phys. Rev. {\bf D 33}, 1314 (1986).
\bibitem{bjorken} 	J. D. Bjorken, Phys. Rev. {\bf D27}, (1983) 140.
\bibitem{bertsch}	  G. F. Bertsch, Nucl. Phys. {\bf A498} (1989) 173c
\bibitem{lund}		B. Andersson et al, Phys. Rep. {\bf 97} (1983) 33
\bibitem{nr}		T. Cs\"org\H o, B. L\"orstad and J. Zim\'anyi,
			"Quantum Statistical Correlations for Slowly
			Expanding Systems", preprint LUNFD6/(NFFL-7084) 1994,
			Physics Letters B in press.
\end{thebibliography}
\end{document}